\documentclass[twocolumn,showpacs,preprintnumbers,superscriptaddress,amsmath,%
amssymb,nofootinbib]{revtex4}

\input psfig.sty

\usepackage{graphicx}
\usepackage{dcolumn}
\usepackage{bm}

\newcommand{\be}{\begin{equation}}
\newcommand{\ee}{\end{equation}}

\begin{document}

\title{Coupled Quintessence and Vacuum Decay}
\author{F. E. M. Costa} \email{ernandes@on.br}

\author{J. S. Alcaniz}\email{alcaniz@on.br}

\affiliation{Departamento de Astronomia, Observat\'orio Nacional,
20921-400, Rio de Janeiro -- RJ, Brasil}

\author{J. M. F. Maia} \email{jmaia@cnpq.br}

\affiliation{COAIE/CNPq, SEPN 509, BL A, 70750-901, Bras\'{\i}lia - DF, Brasil}

\date{\today}

\begin{abstract}
We discuss observational consequences of a class of cosmological models characterized by the dilution of pressureless matter attenuated with respect to the usual $a^{-3}$ scaling due to the decay of vacuum energy. We carry out a joint statistical analysis of observational data from the new \emph{gold} sample of 182 SNe Ia, recent estimates of the CMB shift parameter, and BAO measurements from the SDSS to show that such models favor the decay of vacuum only into the dark matter sector, and that the separately conserved baryons cannot be neglected. In order to explore ways to more fundamentally motivated these models, we also derive a coupled scalar field version for this general class of vacuum decay scenarios. 

\end{abstract}


\maketitle

\section{Introduction}
In the past decade, an impressive convergence of observational results led to a resurgence of interest in a universe dominated by a relic cosmological constant $\Lambda$ (see, e.g., \cite{krauss}). In general grounds, the main motivation behind this idea is the wide belief that the presence of an unclustered component, such as the vacuum energy, not only would explain the observed late-time acceleration of the Universe but would also reconcile the inflationary flatness prediction ($\Omega_{\rm{Total}} \simeq 1$) with the current clustering estimates that point systematically to $\Omega_m \simeq 0.2 - 0.3$ \cite{omega, wmest}.

On the other hand, it is also well known that the same welcome properties that make $\Lambda$CDM scenarios our best description of the observed Universe also result in an unsettled situation in the Particle Physics/Cosmology interface, in which the cosmological upper bound ($\rho_{\Lambda} \lesssim 10^{-47}$ ${\rm{GeV}}^4$) differs from theoretical expectations ($\rho_{\Lambda} \sim 10^{71}$ ${\rm{GeV}}^4$) by more than 100 orders of magnitude \cite{weinberg}. In this regard and based only on phenomenological grounds, an attempt at alleviating such a problem was proposed almost two decades ago, in which $\Lambda$ is allowed to vary \cite{ozer}. Afterward, a considerable number of models with different decay laws for the variation of the cosmological term was investigated in both theoretical \cite{lambdat0} and observational aspects \cite{lambdat1, saulo}. Although constituting a possible way to address the above problem, the usual critique to the so-called  $\Lambda$(t)CDM scenarios is that, in the absence of a natural guidance from fundamental physics, one needs to specify a phenomenological time-dependence for $\Lambda$ in order to establish a model and study their observational and theoretical implications.

In this concern, a still phenomenological but interesting step toward a more realistic decay law was recently discussed by Wang and Meng in Ref.~\cite{wm}.
Instead of the traditional approach specifying a decay law for $\Lambda(t)$ and studying its consequences, they deduced a new one from a simple argument
about the effect of the vacuum decay on the dark matter expansion rate.  Such a decay law is similar to the one originally obtained in Ref.~\cite{shapiro1} from arguments based on renormalization group and is capable of reconciling $\Lambda$(t)CDM models with an initially decelerated and late time accelerating universe, as indicated by current SNe Ia observations \cite{snls,Riess2006}. Additionally, the authors assumed that baryons have a negligible role to play in the late time evolution of the model and supposed that vacuum decays only into dark matter.

\begin{figure*}[t]
\centerline{\psfig{figure=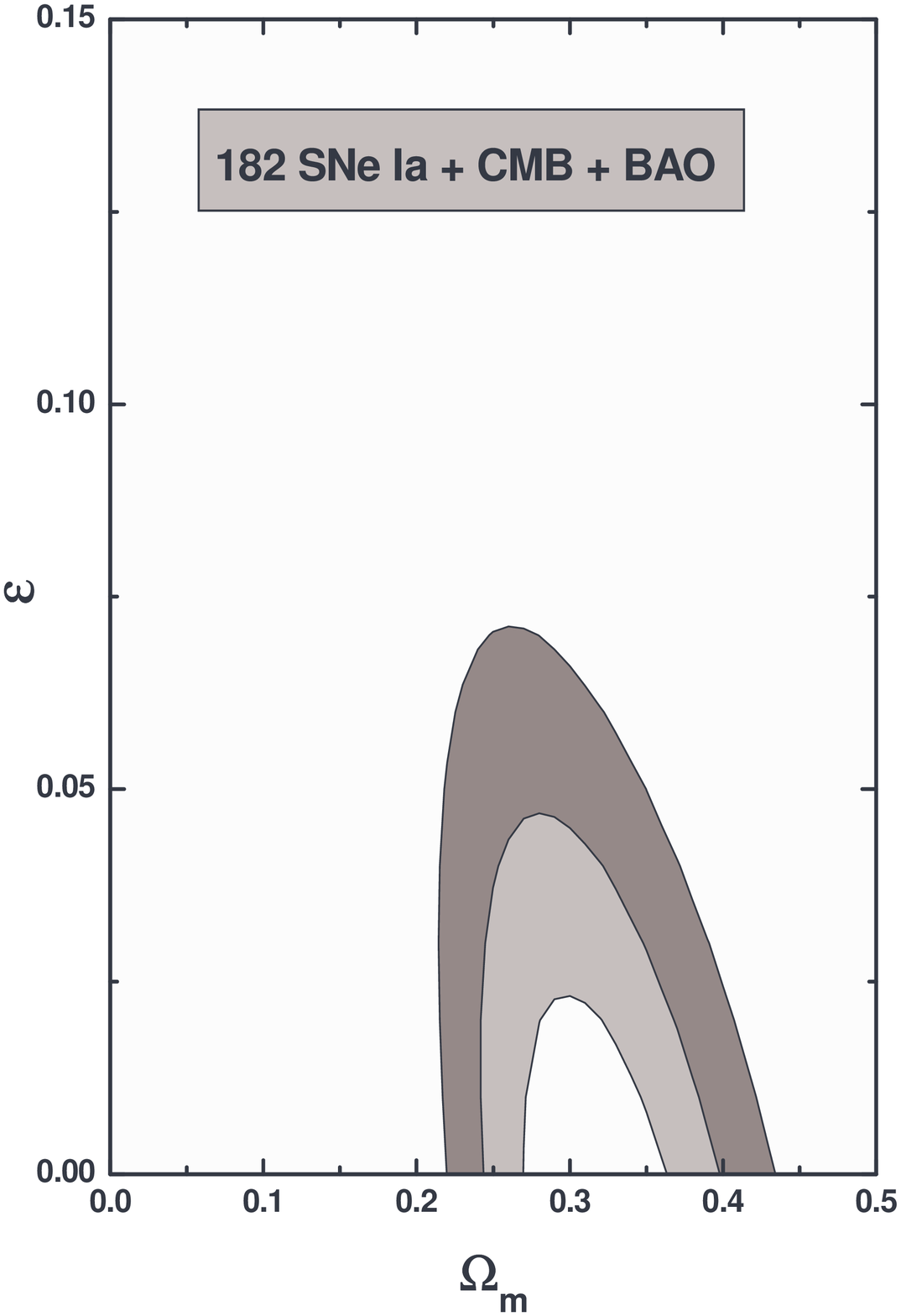,width=3.5truein,height=2.55truein,angle=0}
\psfig{figure=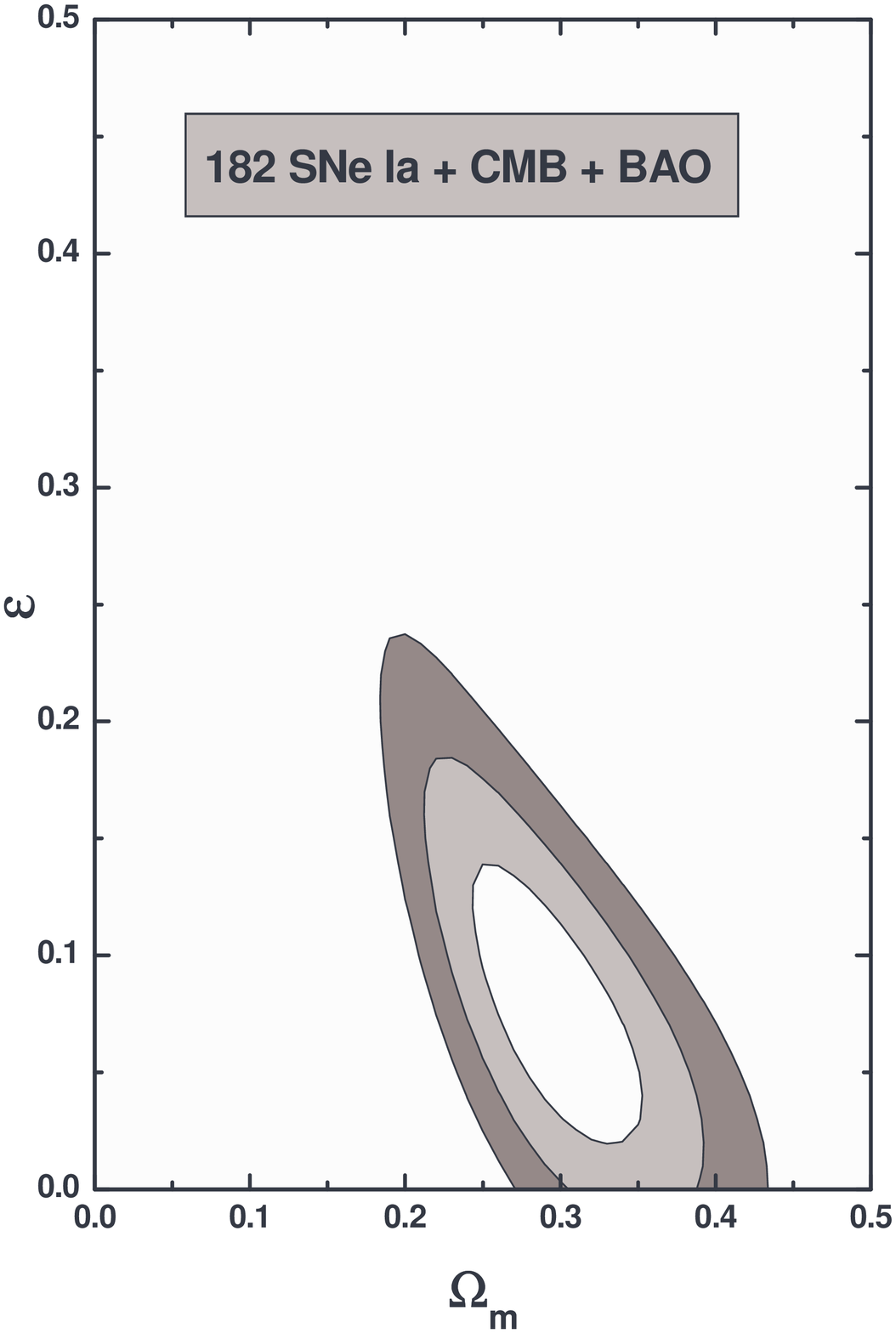,width=3.5truein,height=2.55truein,angle=0}
\hskip 0.1in} \caption{The results of our statistical analyses. In both Panels, the contours correspond to confidence regions of 68.3\%, 95.4\% and 99.7\%. {\bf{Left:}} The plane $\Omega_m - \epsilon$ when the separately-conserved contribution of the baryons to Eq. (\ref{friedmann}), $0.044(1+z)^3$, is either neglected or included as part of the vacuum decay process. The best-fit for this analysis is the current standard $\Lambda$CDM scenario, i.e., $\epsilon = 0$. {\bf{Right:}} The same as in previous Panel when the baryonic content correspond to 4.4\% of the critical energy density. For this analysis, we find $\epsilon \leq 0.13$ and $0.25 \leq \Omega_m \leq 0.35$ at 95.4\% (c.l.).}
\end{figure*}

The qualitative argument used in Ref.~\cite{wm} (see also, \cite{js,js1}) is the following: if vacuum is decaying into dark matter particles, the energy density of this latter
component will dilute more slowly compared to its standard evolution, $\rho_m \propto a^{-3}$, where $a$ is the cosmological scale factor. Thus, the deviation from
the standard dilution is characterized by a positive constant $\epsilon$, such that\footnote{As discussed in Ref.~\cite{js}, thermodynamic arguments applied o the
this process of vacuum decay, more specifically, the second law ($\dot{S}\geq 0$), restrict the parameter $\epsilon$ to be positive.}
\begin{equation}
\label{energyCDM}
\rho_m=\rho_{mo} a^{-3 + \epsilon}.
\end{equation}
This is the simplest possible way of stating the fact that the dark matter dilution is attenuated by the existence of another component of the cosmic constituents decaying into this sector. By introducing the above result into the balance equation for the dark matter particles and $\Lambda$, i.e.,
\begin{equation} \label{ec}
\dot \rho_m + 3\frac{\dot a}{a}\rho_m \ = - \dot \rho_v\;,
\end{equation}
we obtain \cite{wm} 
\begin{equation}\label{decayv}
\rho_{v} =  \tilde\rho_{vo} + \frac{\epsilon \rho_{m0}}{3 - \epsilon}a^{-3 + \epsilon}\;.
\end{equation}
In the above equations, a dot and the subscript 0 denote, respectively, derivative with respect to time and the current values of the parameters, $\rho_{v} = \Lambda/8\pi G$, $\rho_{mo}$ is the dark matter energy density and $ \tilde\rho_{vo}$ is an integration constant. As emphasized in Ref.~\cite{wm}, such a decay law seems to be the most general one, as it has many of the previous phenomenological attempts as particular cases\footnote{It is possible to show that even for  $\Lambda$(t)CDM scenarios whose deviation from the standard evolution of the dark matter energy density is more general than the one given by Eq. (\ref{energyCDM}), e.g., $\rho_m=\rho_{mo} a^{-3} + \rho_{mo} a^{-3 + \epsilon}$ \cite{saulo}, the vacuum decay law of Eq. (\ref{decayv}) remains valid.}. 

Our aim in this paper is twofold: first, to test the viability of this comprehensive class of decaying $\Lambda$ scenarios in light of the latest cosmological observations for an even more general class of models by including baryons in the analysis, as they were not taken into account by the authors of \cite{wm}. To this end, we use the recent type Ia supernovae (SNe Ia) measurements, as given by  the High-$z$ Supernovae Team \cite{Riess2006}, the baryon acoustic oscillations (BAO) measurement from the Sloan Digital Sky Survey (SDSS) \cite{sdss6}, and the shift parameter from the three-year Wilkinson Microwave Anisotropy Probe (WMAP) data \cite{wmap3}.  As we shall show, $\Lambda({\rm{t}})$CDM plus baryons is fitted for a wider range of values of $\epsilon$ if compared to the case without baryons. Second, as a step towards fundamental physics motivation for the attenuated dilution of dark matter, we also discuss here how this class of $\Lambda({\rm{t}})$ cosmologies can be interpreted in terms of a classical scalar field. However, differently from usual scalar field cosmologies, in this case the quintessence field must be coupled to dark matter in order to account for the dilution. Such ``coupled quintessence'' models have been largely considered in the literature to address some fine-tuning issues of the usual quintessence proposals \cite{coupled5}. 

We organized this paper as follows. In Sec. II we investigate current observational constraints on the general class of $\Lambda$(t)CDM models driven by the decay law of Eq. (\ref{decayv}). We show that the best concordance with the data is obtained when the baryonic component is assumed to be a separately-conserved one. We also briefly discuss the age of the Universe in the context of these models. The scalar field version for this class of scenarios is derived in Sec. III. We summarize and discuss our main conclusions in Sec. IV.

\section{Basic equations and Observational Aspects}

Let us first consider a homogeneous, isotropic, spatially flat universe described by the Friedmann-Robertson-Walker line element. In such a background, the
Einstein field equations are given by
\begin{subequations}
\begin{equation}
\label{fried}
8\pi G \rho + \Lambda = 3H^2\;,
\end{equation}
and
\begin{equation}
\label{friedp}
8\pi G p - \Lambda =-2\dot{H} - 3H^2\;.
\end{equation}
\end{subequations}
By combining Eqs. (\ref{energyCDM}) and (\ref{decayv}) with Eq. (\ref{fried}), we can rewrite the Hubble expansion as
\begin{equation}
\label{friedmann}
{\cal{H}} = \left[ \Omega_b(1 + z)^3 + \frac{3\Omega_m}{3 - \epsilon}a^{-3 + \epsilon} + \tilde{\Omega}_{vo}\right]^{1/2}\;,
\end{equation}
where ${\cal{H}}={{H}}/{H_o}$, $\Omega_m$ stands for the current matter density parameter, $\tilde{\Omega}_{vo}$ is the density parameter associated with the residual part of vacuum energy density, i.e., $\tilde\rho_{vo}$, and we have set $a_0 = 1$. Note that, if the baryonic content is either neglected or included in the vacuum decay process, the baryon density contribution is automatically set to be 0. If the latter case is considered, $\Omega_m$ will represent both the baryonic and dark matter density parameters.

\subsection{Observational constraints}

To test the viability of the cosmological scenario described above, we study in this Section observational limits on the parametric space $\Omega_m - \epsilon$. We perform a joint statistical analysis involving four complementary sets of observations, namely, the so-called new \emph{gold} sample of 182 SNe Ia, recently published by Riess et al.~\cite{Riess2006}, the current estimate of the CMB shift parameter given by
\begin{equation}
R \equiv \Omega_m^{1/2}\Gamma(z_{CMB}) = 1.70 \pm 0.03\;,
\end{equation}
from WMAP, CBI, and ACBAR~\cite{wmap3}, where $\Gamma(z)$ is the dimensionless comoving distance and $z_{CMB} = 1089$, and the BAO measurements from the SDSS luminous red galaxies~\cite{sdss6},
\begin{eqnarray}
d_V(z_{*})  =  \left[\Gamma^{2}(z_{*})\frac{cz_{*}}{H(z_{*})}\right]^{1/3} = 1.300 \pm 0.088 \mbox{Gpc}\;,
\end{eqnarray}
obtained in Ref. \cite{teg} from power spectrum estimates, where $z_* = 0.35$ (for more  details on the statistical analysis we refer the reader to Ref.~\cite{refs}). 

In Figs. (3a) and (3b) we show the results of our statistical analyses. Confidence regions (68.3\% and 95.4\%) in the plane $\Omega_m - \epsilon$ are shown for the particular combination of observational data described above. Fig. (3a) corresponds to the case discussed so far, i.e., when the separately-conserved contribution of the baryons to Eq. (\ref{friedmann}), $\Omega_b(1+z)^3$, is either neglected or included as part of the vacuum decay process. Note that  the limits on the parameter $\epsilon$ are very restrictive with the best-fit model corresponding to the standard $\Lambda$CDM dynamics, i.e., $\epsilon = 0$. At 95.4\% c.l., we find for this analysis $\epsilon \leq 0.03$ and $\Omega_m = 0.31^{+0.07}_{-0.06}$.

The effect of 4.4\% of a separately-conserved baryonic contribution on our statistical analysis is displayed in Fig. (3b).  For this analysis, the best-fit values for the decaying $\Lambda$ and the matter density parameters are, respectively, $\epsilon = 0.08$ and $\Omega_m = 0.29$, which correspond to a 9.7$h^{-1}$-Gyr-old, accelerating universe with a deceleration parameter $q_0 = -0.49$ and acceleration redshift $z_a = 0.62$. At 95.4\% c.l., we also find $\epsilon \leq 0.13$. The matter density parameter in this case lies in the interval $0.25 \leq \Omega_m \leq 0.35$, which is in good agreement with current clustering estimates from the relative peculiar velocity measurements for pairs of galaxies \cite{wmest} to CMB observations \cite{wmap3}. 

\subsection{Age of the Universe}

For the sake of completeness, we also derived the age-redshift relation for this class of $\Lambda {\rm{(t)}}$CDM models , i.e.,
\begin{equation}
t(a) = -\frac{H_0^{-1}}{\sqrt{3{\rm{A}}(3 - \epsilon) \Omega_m}}\ln{\left[\frac{\sqrt{a^{-3 + \epsilon} + {\rm{A}}} - \sqrt{{\rm{A}}}}{\sqrt{a^{-3 + \epsilon} + {\rm{A}}} + \sqrt{{\rm{A}}}}\right]}
\end{equation}
where $a = (1 + z)^{-1}$ and ${\rm{A}} = \frac{3 - \epsilon}{3 \Omega_m} - 1$. Fig. (2) shows the age of the Universe as a function of the redshift for some selected values of the  parameter $\epsilon$, $\Omega_m = 0.27$, as given by current CMB measurements \cite{wmap3} and $H_0 = 72$ ${\rm{km}/s/Mpc}$, as provided by the final results of the HST {\it key} project \cite{hst}. Note that, at a given redshift,  the larger the value of the decaying parameter $\epsilon$, the larger the age of the Universe that is predicted by these $\Lambda {\rm{(t)}}$CDM models. The standard $\Lambda$CDM model, formally equivalent to the case $\epsilon = 0$, is also shown (solid line) for the sake of comparison.

\begin{figure}
\centerline{\psfig{figure=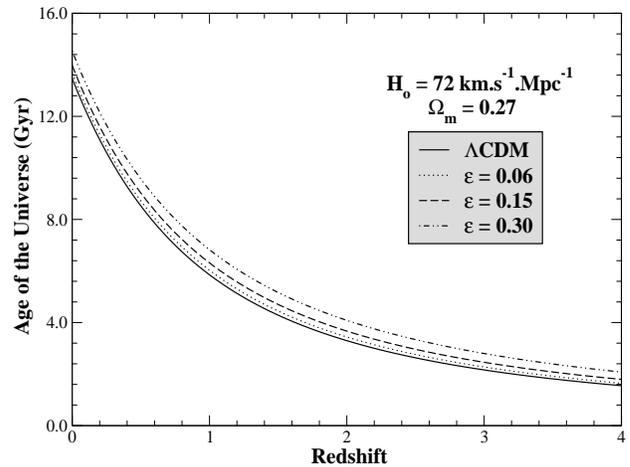,width=3.4truein,height=2.7truein,angle=-90}
\hskip 0.1in} \caption{Age of the Universe as a function of the redshift for seleceted values of $\epsilon$, $\Omega_m = 0.27$ and $H_0 = 72$ ${\rm{km}/s/Mpc}$. As discussed in the text, at a given redshift,  the larger the value of the decaying parameter $\epsilon$, the larger the age of the Universe that is predicted by the $\Lambda {\rm{(t)}}$CDM model.}
\end{figure}

\section{Scalar field description}
In order to find the scalar field counterpart for the flat decaying $\Lambda$ model considered here (that shares the same dynamics and temperature evolution law), we use the procedure originally proposed in Ref.~\cite{jmaia} (see also \cite{jmaia1} for a general description with arbitrary curvature).

First, let us define the parameter
\begin{equation}
\gamma_* = - \frac{2\dot{H}}{3H^2} =  (1 - \frac{\Lambda}{3H^2})\;,
\end{equation}
which is just another way of writing the field equations (\ref{fried}) and (\ref{friedp}). Thus, any cosmology dynamically equivalent to the  $\Lambda {\rm{(t)}}$CDM model considered here must have the same $\gamma_*$. By combining the above equation with Eqs. (\ref{energyCDM}) and (\ref{decayv}) we obtain
\begin{equation}
\gamma_* = \frac{3 - \epsilon}{3}\frac{a^{-3 + \epsilon}}{{\rm{B}} + a^{-3 + \epsilon}}\;,
\end{equation}
where ${\rm{B}} = \frac{3 - \epsilon}{3}\frac{\tilde{\rho}_{vo}}{\rho_{m0}}$. Following standard lines, we replace the vacuum energy density and pressure in Eqs. (\ref{fried}) and (\ref{friedp}) by the corresponding scalar field expressions, i.e.,
\begin{equation}
\rho_v = \frac{\Lambda}{8\pi G} \rightarrow \rho_{\phi} \quad \mbox{and} \quad p_v = -\frac{\Lambda}{8\pi G} \rightarrow p_{\phi}\;,
\end{equation}
where $\rho_{\phi} = \dot{\phi}^2/2 + V(\phi)$ and $p_{\phi} = \dot{\phi}^2/2 - V(\phi)$ are, respectively, the energy density and pressure associated with the coupled scalar field $\phi$ whose potential is $V(\phi)$.

By defining a new parameter
$x \equiv {\dot\phi^2}/{(\dot\phi^2 + \rho_m)}$ with $0 \leq x \leq 1$,
we can manipulate the above equations to obtain (see \cite{jmaia} for more details)
\begin{subequations}
\begin{equation}\label{eq:dotphi}
\dot{\phi}^2 =  {3H^2 \over 8\pi G} \gamma_* x\;,
\end{equation}
and
\begin{equation}\label{eq:V}
V(\phi) = {3H^2 \over 8\pi G} \left[1 - \gamma_* \left(1-{x\over
2}\right)\right]\;,
\end{equation}
\end{subequations}
which link directly the field and its potential with the related quantities of the dynamical $\Lambda{\rm{(t)}}$ case. From Eq. (\ref{eq:dotphi}), one can show that in terms of the scale factor $a$ the field $\phi$ is given by
\begin{eqnarray} \label{ph}
\phi & & = \sqrt{\frac{3}{8\pi G}}\int{\sqrt{\gamma_* x}\frac{da}{a}}  \\ \nonumber & &
= {\rm{C}} \times \ln\left({  \sqrt{{\rm{B}} + a^{-3 + \epsilon}} + \sqrt{a^{-3 + \epsilon}}}\right)\;,
\end{eqnarray}
where ${\rm{C}} = -[\frac{x}{2\pi G(3 - \epsilon)}]^{1/2}$. 
Note that, to derive the above expression, we have assumed $x$ to be a constant, which is equivalent to impose the condition that the scalar field version mimics exactly the particle production rate of its decaying $\Lambda({\rm{t}})$ counterpart \cite{jmaia}. Now, by inverting the above equation and inserting $a(\phi)$ into Eq. (\ref{eq:V}), the potential $V(\phi)$ is readily obtained, i.e.\footnote{Double exponential potentials of the type (\ref{vphi}) have been considered in the literature as viable examples of quintessence scenarios (see, e.g, \cite{dp}). As discussed in Ref.~\cite{dp1}, a field potential given by the sum of two exponential terms is also motivated by dimensional reduction in M-theory with interesting implications for the late-time accelerating behavior of the cosmic expansion.},
\begin{equation} \label{vphi}
V( \phi )  = \tilde{\rho}_{vo} + {\rm{D}}\left[e^{\frac{2\phi}{{\rm{C}}}} + {\rm{B}}^2e^{-\frac{2\phi}{{\rm{C}}}} - 2{\rm{B}} \right]
\end{equation}
where ${\rm{D}} = \rho_{m0}\frac{(3 - \epsilon)x + 2\epsilon}{8(3 - \epsilon)}$. As one may check, as $a \rightarrow \infty$, $\phi = {\rm{C}} \times \ln(\sqrt{{\rm{B}}}) = \rm{const.}$, so that the associated energy density of the field [$\rho_{\phi} = \dot{\phi}^2/2 + V(\phi)$] is fully specified by its potential. Thus, by substituting this latter value for the field into Eq. (\ref{vphi}) we obtain
\begin{equation}
\rho_{\phi} \equiv V(\phi) = \tilde{\rho}_{vo}\;, \quad \quad (a \rightarrow \infty)
\end{equation}
which is exactly the asymptotic limit for the vacuum energy density $\rho_v$, as given in Eq. (\ref{decayv}).

\begin{figure}[t]
\centerline{\psfig{figure=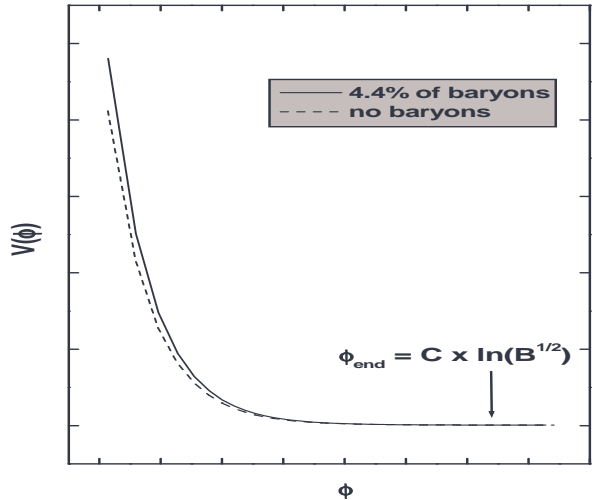,width=3.4truein,height=2.8truein,angle=0}
\hskip 0.1in} \caption{The potential $V(\phi)$ as a function of the field [Eq. (\ref{vphi})]. Note that the general case in which the baryons are included as a separately-conserved component corresponding to 4.4\% of the critical density represents only a small shift relative to the case without baryons (dashed line). The arrow indicates the final value of the field, $\phi_{\rm{end}} = {\rm{C}} \times \ln(\sqrt{{\rm{B}}})$, when $a \rightarrow \infty$.}
\end{figure}

Finally, an important aspect worth emphasizing is that to derive the potential (\ref{vphi}) we have either neglected or included the baryonic content as part of the vacuum decay process. This has been done because in the most general case, when the baryons are included as a separately-conserved component, Eqs. (\ref{eq:dotphi}) and (\ref{eq:V}) cannot be analitically solved. Numerically, however, we can show that the effect of $\simeq 4.4\%$ of baryons on the new potential $V(\phi)$ is only a small shift relative to the previous case, so that our general potential belongs to the same class of potentials given in Eq. (\ref{vphi}) and shown in Fig. (3).

\section{Final Remarks}


In this paper, we have studied observational and theoretical consequences of an alternative mechanism of cosmic acceleration based on a general class of decaying vacuum scenarios whose decaying law is given by Eq. (\ref{decayv}). Differently from other recent phenomenological proposals discussed in the literature, such a decaying law is deduced only from the effect of a time-varying $\Lambda$ term on the dark matter evolution with the scale factor \cite{wm, js, js1}. From the observational viewpoint, we have shown that a process of vacuum decay only into cold dark matter particles (with the baryonic contribution separately conserved) is favoured over the case in which the baryons also participate of the process (or are neglected). In the former case, for instance, we have found $\epsilon \leq 0.13$ at 2$\sigma$ level, which constitutes a small but measurable deviation from the standard $\Lambda$CDM dynamics (formally equivalent to the case $\epsilon = 0$).

We have also discussed a scalar field description for these  cosmologies and found that this class of $\Lambda$(t)CDM models is identified with a coupled quintessence field with the potential of Eq. (\ref{vphi}). Since this class of $\Lambda$(t)CDM scenarios is quite general (having many of the previous proposals as a particular case), we argue that a coupled quintessence field model whose potential is given by the double exponential function of Eq. (\ref{vphi}) is dinamically equivalent to a large number of decaying vacuum scenarios previously discussed in the literature.

\begin{acknowledgments}
JMFM would like to thank the hospitality of the Departamento de Astronomia of Observat\'orio Nacional/MCT where part of this work was developed. FEMC acknowledges financial support from CAPES. JSA is supported by CNPq (No. 307860/2004-3 and  No. 475835/2004-2) and by FAPERJ (No. E-26/171.251/2004).
\end{acknowledgments}

\end{document}